# Development of helium turbine loss model based on knowledge transfer with Neural Network and its application on aerodynamic design

Changxing Liu [a, c], Zhengping Zou [a, c, *], Pengcheng Xu [a, b], Yifan Wang [a, c]

a. National Key Laboratory of Science and Technology on Aero-Engine Aero-thermodynamics, Beijing 100191, China

b. School of Energy & Power Engineering, Beihang University, Beijing 100191, China

c. Research Institute of Aero-Engine, Beihang University, Beijing 102206, China

**Abstract**

Helium turbines are widely used in the Closed Brayton Cycle for power generation and aerospace applications. The primary concerns of designing highly loaded helium turbines include choosing between conventional and contra-rotating designs and the guidelines for selecting design parameters. A loss model serving as an evaluation means is the key to addressing this issue. Due to the property disparities between helium and air, turbines utilizing either as working fluid experience distinct loss mechanisms. Consequently, directly applying gas turbine experience to the design of helium turbines leads to inherent inaccuracies. A helium turbine loss model is developed by combining knowledge transfer and the Neural Network method to accurately predict performance at design and off-design points. By utilizing the loss model, design parameter selection guidelines for helium turbines are obtained. A comparative analysis is conducted of conventional and contra-rotating helium turbine designs. Results show that the prediction errors of the loss model are below 0.5% at over 90% of test samples, surpassing the accuracy achieved by the gas turbine loss model. Design parameter selection guidelines for helium turbines differ significantly from those based on gas turbine experience. The contra-rotating helium turbine design exhibits advantages in size, weight, and aerodynamic performance.

\* Corresponding author.

*E-mail address:* zouzhengping@buaa.edu.cn (Z. Zou)

# 1. Introduction

The high energy density of helium and its low viscosity in a supercritical state renders helium Closed Brayton Cycle a vital component in various energy systems. Within these systems, helium turbines are crucial in influencing overall efficiency. Compact and high-efficiency helium turbines are essential for hypersonic precooled engines that are sensitive in size and weight [1–3]. To fulfill the requirements of nuclear power generation systems for high power output and economic efficiency, helium turbines are considered critical components for energy conversion in high temperature gas cooled reactors[4–7]. The low thermal efficiency due to the extremely low temperature makes helium turbines the most critical component of the large-scale helium cryogenic system, which plays a pivotal role in superconducting systems, high-energy physics, and space exploration [8–10].

The low molecular weight and high specific heat capacity of helium result in a great number of stages in helium turbines. To achieve a more compact energy system, reducing the number of stages in helium turbines is necessary, which means increasing the load carried by each stage. Consequently, highly loaded and high-efficiency helium turbine design is of great significance. The major problem for highly loaded helium turbine designing is the choice between conventional or contra-rotating design and design parameter selection guidelines, which requires a high precision loss model applicable to helium turbines.

Many turbine loss models have been developed over the years. Smith[11] used a large amount of turbine performance data to create a simple correlation chart of the load coefficient and flow coefficient with efficiency. Ainley and Mathieson[12] proposed a semi-empirical correlation between turbine geometry parameters and losses based on turbine experiment data and proposed a semi-empirical correlation between turbine geometry parameters and losses based on turbine experiment data. With the advances in aerodynamic analysis and the accumulation of turbine data, this correlation was revised by Dunham and Came[13] and further developed by Kacker and Okapuu[14] into a classic turbine loss model. Denton[15] derived analytical expressions for various losses in turbines based on the internal loss mechanism of the turbine. Recently, with the improvement in the understanding of turbine design and the application of different mathematical methods, loss models that take more factors into account have been proposed to achieve higher accuracy in loss prediction[16–23]. Based on extensive research on loss models, as mentioned above, numerous scholars have made significant advancements in developing methods for designing helium turbines. Braembussche et al.[24] proposed helium turbine low-dimensional parameter selection guidelines based on the gas turbine Smith chart, leading to the design of a multi-stage helium turbine for high temperature gas-cooled nuclear reactors. Tournier and Mohamed[25] developed design models for multi-stage axial turbines and compressors based on previous studies of gas turbines and applied them to the design and performance evaluation of helium turbines. Li et al. [10] presented a one-dimensional mean line design and optimization approach for helium turbines within helium cryogenic systems.

In general, the contra-rotating design can effectively increase the load per stage and reduce the overall size. Early researchers claimed that the improvement in aerodynamic efficiency from contra-rotating turbines could be as high as 4%[26]. Louis[27] conducted a detailed study of several types of contra-rotating turbines and concluded that contra-rotating turbines have advantages over conventional turbines in terms of efficiency, size, and output work and established a Smith chart for contra-rotating turbines. To compare conventional and contra-rotating turbines, Waldren et al.[28] propose the relative parameter and conclude that with the same midspan radius, the contra-rotating turbine has four times the power output per stage, but the relative Mach number doubles, which will introduce complexities to the design. The speed of sound in a helium turbine is three times higher than in a gas turbine, resulting in a relatively low Mach number. Based on this, Varvill et al.[29] conclude that the contra-rotating helium turbine has advantages in the hypersonic precooled engines and design a four-stage contra-rotating turbine that achieves 94.8% polynomial efficiency.

However, the existing research on helium turbine design methods largely relies on gas turbine loss models and design experience. The disparities between helium and gas turbines cast doubts about the reliability of these models and methods. Furthermore, existing studies have predominantly focused on helium turbines with low loading. These raise a significant challenge for the design of highly loaded helium turbines. In response to this quandary, this paper aims to establish a high-precision helium turbine loss model, derive low-dimensional parameter selection guidelines, and identify the distinguishing features of conventional and contra-rotating helium turbines.

The establishment of a high-precision helium turbine loss model relies fundamentally on the existence of a comprehensive and accurate helium turbine database. Because of the high cost and the danger of high-pressure helium leaks, similar methods for different working fluids have been developed for the experimental study of helium turbomachinery. Robert and Sjolander[30] derived the effect of isentropic exponent on the performance conversion of turbomachinery with different working fluids. David et al.[31] proposed the effect of isentropic exponent on transonic turbine loss. Ding[32] and Chen[33] proposed similar methods for turbomachinery with different working fluids. However, inherent errors in the similarity process do not meet the requirements for constructing a high-precision helium turbine database. Computational Fluid Dynamics (CFD) method, on the other hand, provides a more convenient approach to attain the aerodynamic performance for helium turbines. Therefore, the CFD method is exceptionally suitable for the database.

An appropriate modeling method is crucial for constructing a high-precision loss model in the context of limited data availability. With the development of computer science, machine learning is increasingly employed in the turbomachinery field. An early application is an approximate model proposed by Pierret[34] using Neural Networks combined with a CFD solver for the design and optimization process of turbine blades, effectively shortening the design period. Son[35] developed an off-design model for supercritical $CO_2$ turbomachinery based on deep Neural Networks, significantly reducing the prediction error compared to conventional models. Xu[36] established a parameter prediction model using a deep Neural Network, which improved the accuracy of performance conversion for similar compressors working with different gases. Du[37] proposed a blade end wall profile design and optimization model for steam turbines by employing a convolutional Neural Network. Chen[38] developed a two-dimensional blade profile design and optimization method based on Neural Networks and multi-output Gaussian processes. In recent years, transfer learning has been widely used in various fields. Transfer learning improves the performance of the target model on target domains by transferring the knowledge contained in different but related source domains, which will reduce the dependence on a large amount of target-domain data. In recent years, transfer learning has been widely used in various fields. Transfer learning improves the performance of the target model on target domains by transferring the knowledge contained in different but related source domains, which will reduce the dependence on a large amount of target-domain data[39,40]. The difficulty of obtaining large amounts of data for helium turbines and the well-established research on gas turbine loss models make transfer learning suitable for helium turbine modeling.

In this paper, an extensive database of helium turbines is obtained through fast design and analysis. Knowledge transfer combined with the Neural Network is performed to develop the loss model for helium turbines. Guidelines for the low-dimensional aerodynamic design of helium turbines are investigated based on the developed loss model. Finally, the aerodynamic design of highly loaded conventional and contra-rotating helium turbines is carried out to compare their characteristics.

The rest of this paper is arranged as follows: Section 2 is concerned with the overall framework and the modeling methodology, Section 3 gives the modeling results with an accuracy analysis and low-dimensional design parameter selection guidelines, Section 4 presents the aerodynamic design of the conventional and contra-rotating helium turbines and the comparative analysis, Section 5 provides a summary.

## 2. Methodology

This section comprehensively describes developing a helium turbine loss model employing knowledge transfer and Neural Network. The overall framework is presented, followed by an outline of the knowledge transfer method. Subsequently, the specific structures and training methods for the loss model at the design and off-design points are provided. Lastly, the process for establishing a high-precision helium turbine database with a sufficient range of parameters is described.

**2.1 Overall Framework**

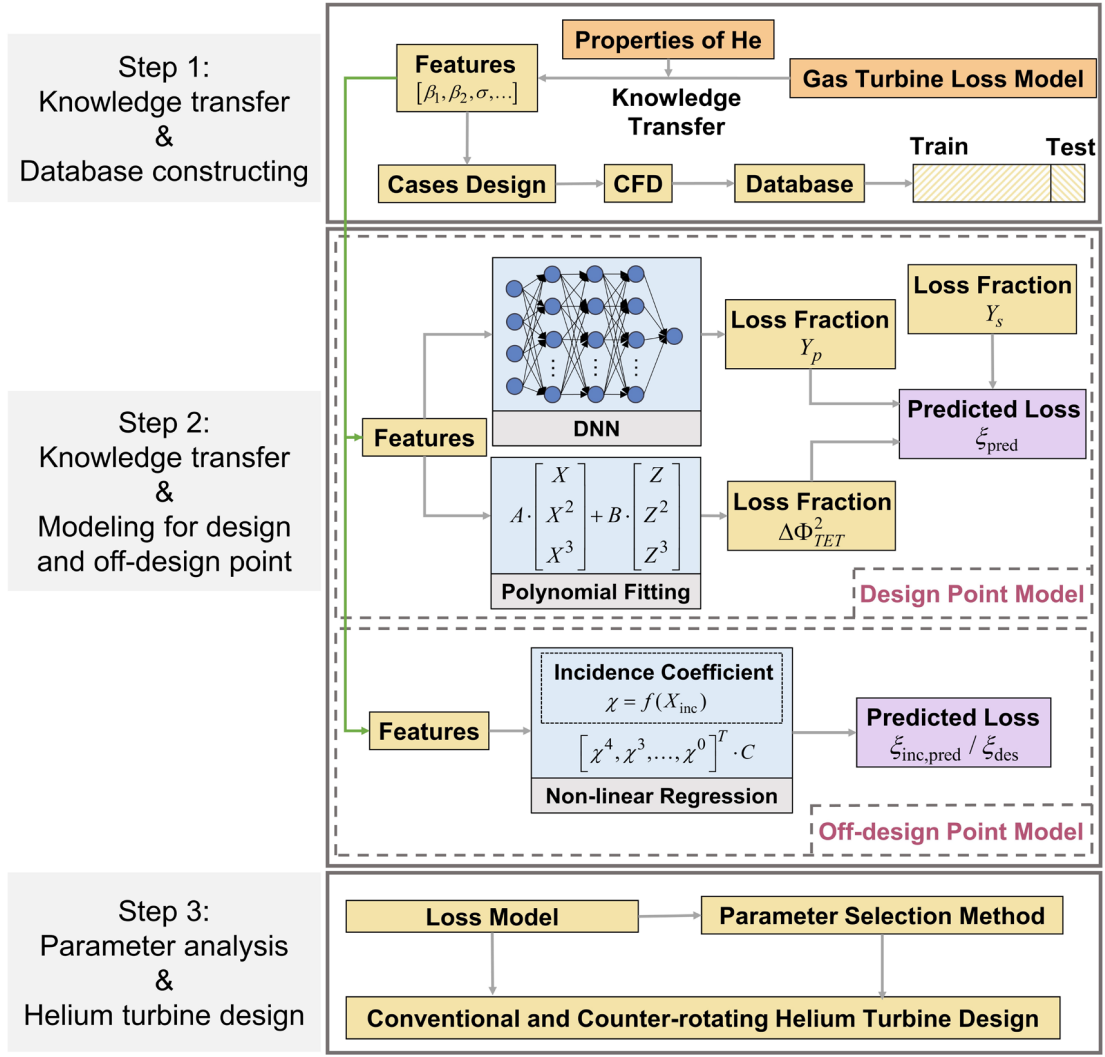

**Fig. 1.** Overall framework of the modeling method

Fig. 1 depicts the three-step framework of the modeling approach. In the first step, knowledge is transferred from the gas turbine model to create a database for the helium turbine, which is crucial for the loss model development. The second step focuses on developing the helium turbine loss model using knowledge transfer and Neural Network methods at design and off-design points. The third step utilizes the established loss models to derive selection guidelines for low-dimensional helium turbine design parameters. These methods are then employed in designing and comparing a conventional and a contra-rotating helium turbine.

A multi-level knowledge transfer approach is utilized in modeling the helium turbine losses. The gas turbine loss model is employed as the source model. The knowledge transfer includes two parts: features of the model and the encompassed data information. Details of the knowledge transfer are illustrated in Fig. 2. By extracting the features from the source model and considering the differences between helium and air, the specific features of the helium turbine loss model are derived, which achieves the transfer of features. The model transfer involves transferring the mathematical correlations from the source model while partially preserving them and introducing additional enhanced model structures, such as Neural Networks. This results in a helium turbine loss model with coefficients to be trained. The data transfer process begins with generating a dataset utilizing the source model, which contains valuable information. Subsequently, the target model undergoes pre-training to facilitate the transfer of the data.

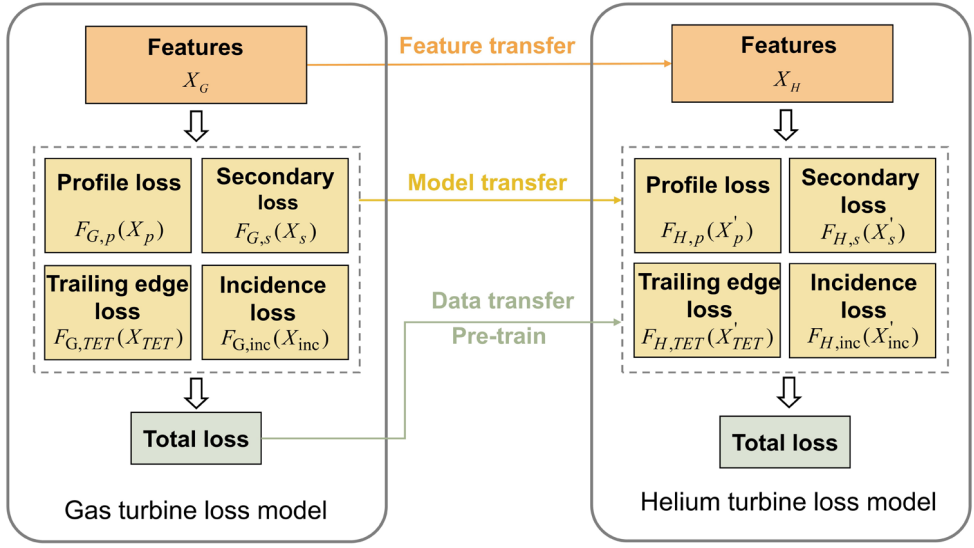

**Fig. 2.** Schematic of the knowledge transfer process

The modeling problem can be formulated as follows:

$$\min_{f_H} \sum_{i=1}^{n_H} L\left[ f_H(X_i) - \xi_i \right] \tag{1}$$

Where the vector $X_i$ is the parameters of the helium turbine, the scalar $\xi_i$ is the loss coefficient from the database, and L is the evaluation function to assess the model error.

**2.2 Knowledge transfer and modeling**

2.2.1 Loss model at the design point

Profile loss:

The profile loss part from the KO model[14] is selected as the source model. The source model encompasses the basic loss term derived from experimental data fitting and correction factors about the Reynolds number, the compressibility, and the shock effects. However, there is a limitation in the basic loss term $Y_{p,AM}$, which is derived from experimental data obtained from axial inlet guide vanes and impulse blades. This correlation is primarily based on blades with an inlet flow angle greater than or equal to zero, posing challenges when applied to blades with a negative inlet flow angle, commonly found in contra-rotating turbines. In such cases, the calculation becomes an external interpolation of the experimental data, leading to significant errors. Furthermore, due to the difference in properties between helium and air, the experimental data from the air may not be directly applicable to helium turbines. As a result, it becomes necessary to refactor the basic loss term.

The Mach number in the helium turbine is relatively low, for the speed of sound in the helium turbine is 3.6 times higher than in the air turbine, thereby allowing the neglect of shock losses. On the other hand, the compressibility of the working fluid and the flow acceleration in the channel cannot be disregarded, emphasizing the need to retain corrections for compressibility and flow acceleration.

Based on the analysis above, the helium turbine profile loss model is described as follows: 1) the features of the model are $[\beta_1, \beta_2, t_{max}/C, \sigma, Re, Ma]$. 2) The basic profile loss term needs to be refactored. 3) neglecting the shock loss term. Consequently, the profile loss correlation can be obtained as Eq (2).

$$Y_p = K_{in} Y_{p,basic} K_p K_{Re} \qquad (2)$$

Where $Y_{p,basic}$ is the basic profile loss term, $K_{in}$ is the correction factor related to the type of the turbine, $K_p$ is the correction term for compressibility, and $K_{Re}$ is the correction for Reynold number.

A Neural Network is employed to establish the correlation between the features and the basic loss. After numerous experiments with hyperparameters, a four-layer Neural Network model is selected with four features, $[\beta_1, \beta_2, t_{max}/C, \sigma]$, as model inputs and the value of the basic loss, $Y_{p,basic}$, as the output. The Neural Network structure is visualized in Fig. 3. Each unit in the Neural Network contains a weight term and an intercept term. The hyperparameter details of the Neural Network are provided in Table 1.

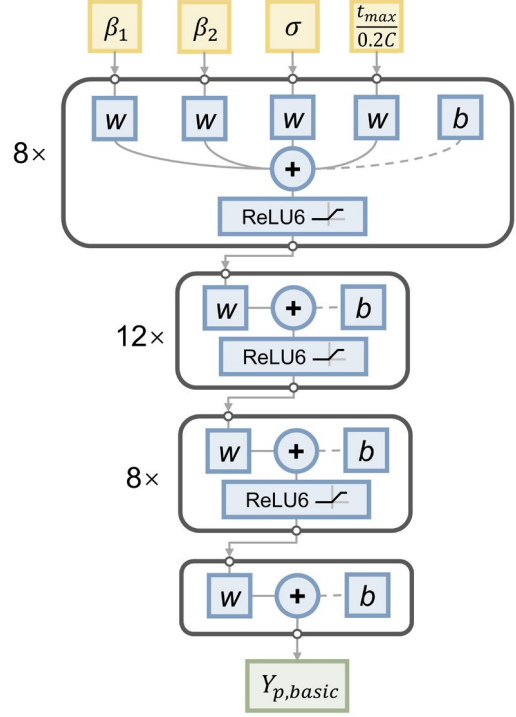

**Fig. 3.** Architecture of the DNN for basic profile loss

**Table 1** Neural Network structure of basic profile loss term

| Parameters | Value |
|---|---|
| Feature dimensions | 4 |
| Hidden layers | 3 |
| Hidden units | 8/12/8 |
| Output dimensions | 1 |
| Activation function | *ReLU6* |

The factor $K_{in}$ and the Reynolds number correction factor proposed by Zhu and Sjolander[41] are employed in this loss model, given as:

$$K_{in} = \begin{cases} 0.825, & \text{for IGVs} \\ 2/3, & \text{for reaction blades} \end{cases} \quad (3)$$

$$K_{Re} = \begin{cases} \left(2\times10^5 / Re_C\right)^{0.575}, & \text{for } Re_C < 2\times10^5 \\ 1.0, & \text{for } Re_C \geq 2\times10^5 \end{cases} \quad (4)$$

The factor $K_p$, accounting for the compressibility, is identical to that introduced by Kacker and Okapuu [14], which is calculated as follows:

$$K_p = 1 - K_2(1 - K_1),$$
$$K_1 \begin{cases} 1 & , \text{for } Ma_2 < 0.2 \\ 1 - 1.25 \times (Ma_2 - 0.2), \text{for } Ma_2 \geq 0.2 \end{cases}, \quad (5)$$
$$K_2 = (Ma_1 / Ma_2)^2.$$

Considering the effect of the end wall secondary flow region, the profile loss is corrected by employing the Spanwise penetration depth proposed in [18].

Trailing edge loss:

The profile loss model mentioned above assumes a zero trailing edge thickness. Hence it is necessary to introduce a trailing edge loss term that accounts for the effects originating from the trailing edge thickness. The trailing edge loss term in the source model faces the same issue as the profile loss term. Consequently, the trailing edge loss term needs to be refactored. The ratio of trailing edge thickness to throat, denoted as $t_{TE}/O$, is employed to demonstrate the effect of flow blockage, and the ratio of inlet and outlet flow angles, $\beta_1/\beta_2$, serves as a representation of the boundary layer at the railing edge, with the exclusion of shock loss. In summary, features of the trailing edge thickness loss model are $[\beta_1/\beta_2, t_{TE}/O]$ and the loss correlation is formulated through the application of a polynomial fitting as follows:

$$\Delta \Phi_{TET}^2 = \sum_{n=1}^{3} \left( a_i - b_i \frac{\beta_1}{\beta_2} \left| \frac{\beta_1}{\beta_2} \right| \right) \left( \frac{t_{TE}}{O} \right)^i \quad (6)$$

where $a_n, b_n (n = 1, 2, 3)$ are the coefficients that need to be determined. The absence of a interpret term is because the loss does not exist when the trailing edge thickness is zero. The term $\beta_1/\beta_2$ indicates the blade type and is multiplied by its absolute value to allow for the application to blades with negative inlet flow angles.

The kinetic energy loss coefficient is converted to a pressure loss coefficient using the following equation:

$$Y_p = \frac{\left\{ 1 - \frac{\kappa-1}{2} Ma_2^2 \times \left( \frac{1}{1-\Delta\Phi^2} - 1 \right) \right\}^{-\frac{k}{k-1}} - 1}{1 - \left( 1 + \frac{\kappa-1}{2} Ma_2^2 \right)^{-\frac{k}{k-1}}} \quad (7)$$

Secondary loss:

The correlation devised by Benner [18,19] is employed to calculate the secondary loss. This model considers not only factors such as blade loading and aspect ratio but also incorporates endwall inlet boundary layer thickness. As a result, it has achieved a higher level of predictive accuracy. The features of the secondary loss model include convergence ratio $\cos\beta_1/\cos\beta_2$, endwall surface area $C_x/(H\cos\beta_2)$, aspect ratio $H/C$, and stagger angle $\gamma$.

Model training method:

Based on the derived characteristics and structure of the helium turbine loss model, the total loss is calculated using Eq (8), which serves as the predicted loss for the design point of the loss model.

$$\xi = \sqrt{1-\Delta\Phi_p^2 - \Delta\Phi_s^2 - \Delta\Phi_{TET}^2} \tag{8}$$

Fig. 4 illustrates the computational chart of the loss model at the design point. The variables in square brackets are the features of the model, and the light blue background indicates the parameters to be trained. The basic profile loss, $Y_{p,basic}$, is computed utilizing the Neural Network and then corrected to obtain the profile loss $Y_p$. The trailing edge loss, $\Delta\Phi_{TE}$, is obtained through a polynomial fitting method. Then the secondary flow loss is calculated. Subsequently, the total velocity loss coefficient is determined by Eq (8). The mean squared error (MSE) loss function is utilized to assess the error, with its calculation expression provided in Eq (9)。

$$MSE = \frac{1}{N}\sum_{j=1}^{N}\left[\xi_j - f(X_j)\right]^2 \tag{9}$$

The components of the model to be trained include a Neural Network in the profile loss part and a fitting polynomial in the trailing edge loss part. The training of the Neural Network employs the Adam optimization algorithm [42]，while the momentum gradient descent optimization algorithm is applied to the fitting polynomial. Pre-train of the model is accomplished separately using data from the source model. Subsequently, every part of the model is simultaneously trained using the helium turbine database. The training process adopts the Mini-batch mode to reduce memory requirements and enhance training efficiency. The Neural Network and trailing edge loss formula are trained with initial learning rates of 0.004 and 0.01, respectively. After every 1200 epochs of training, the learning rates are halved.

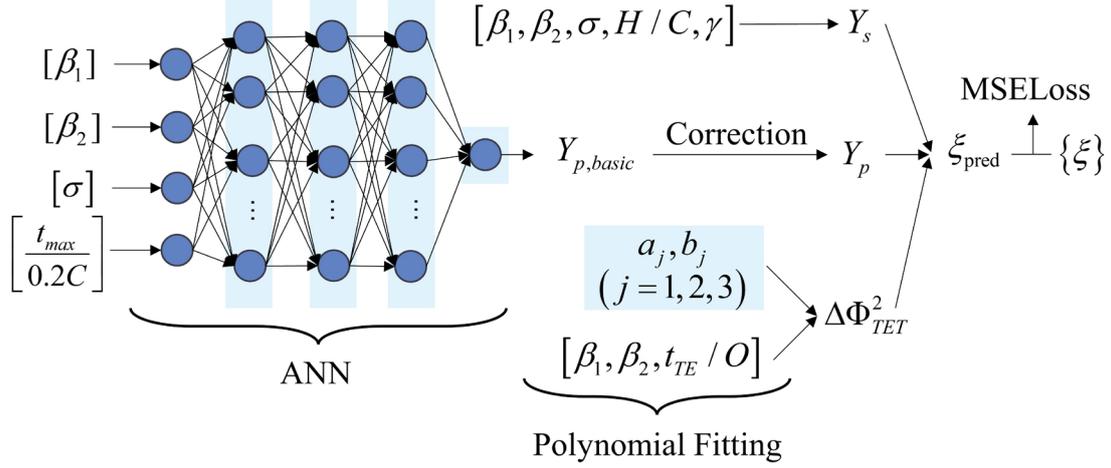

**Fig. 4.** Computation graph of design point loss model

2.2.2 Incidence loss model at the off-design point

The construction of the off-design point loss model consists of two components: converting the features and incidence angle into a dimensionless incidence coefficient and establishing the correlation between the coefficients and the incidence loss. Based on the results of the knowledge transfer, the features are the convergence ratio $\cos\beta_1/\cos\beta_2$, which reflects flow acceleration, the turning angle $(\beta_1+\beta_2)$, which indicates the blade loading, and the relative leading edge thickness $t_{TE}/C$, which represents the leading edge geometry. Utilizing these three features, the incidence angle is converted into a dimensionless coefficient in the form of Eq (10):

$$\chi = \frac{i}{\left(\dfrac{\cos\beta_1}{\cos\beta_2}\right)^{k_1}(\beta_1+\beta_2)\left(\dfrac{t_{TE}}{C}\right)^{k_2}} \tag{10}$$

where $k_1, k_2$ are the coefficients to be determined. The exponent of the turning angle term is fixed to 1 to ensure that the coefficient is dimensionless. The following step establishes the correlation between the incidence coefficient and the incidence loss. A polynomial relationship is constructed as shown in Eq (11):

$$\frac{\xi}{\xi_{des}} = \sum_{n=1}^{m} p_n \chi^n + q \tag{11}$$

When the incidence angle is 0, that is, the same as the design conditions, the loss should be the same as the design point loss, so $q$ is fixed at 1, and $p_n$ are coefficients to be determined.

The training of the incidence loss model requires the simultaneous determination of all the coefficients, which poses a nonlinear fitting problem. To mitigate potential underfitting or overfitting issues, several attempts are conducted then the order of Eq (11) is set to 4, which strikes an optimal balance. The training process of the model consists of pre-training using the data generated from the source model and training using the helium turbine database. The momentum gradient descent is employed as an optimization algorithm. The mini-batch mode is selected to reduce the memory requirement and improve the training efficiency. The learning rate is set to 0.01 and halved after every 500 epochs.

**2.3 Database of helium turbines**

The construction of helium turbine loss models relies on a comprehensive helium turbine database with sufficient parameter ranges. Fig. 5 provides the flowchart of the database construction process. The features obtained through knowledge transfer serve as the variables in the design of the experiment procedure. Subsequently, aerodynamic design and CFD computation of helium turbine cases are conducted to generate datasets. These datasets are then partitioned into train sets and test sets.

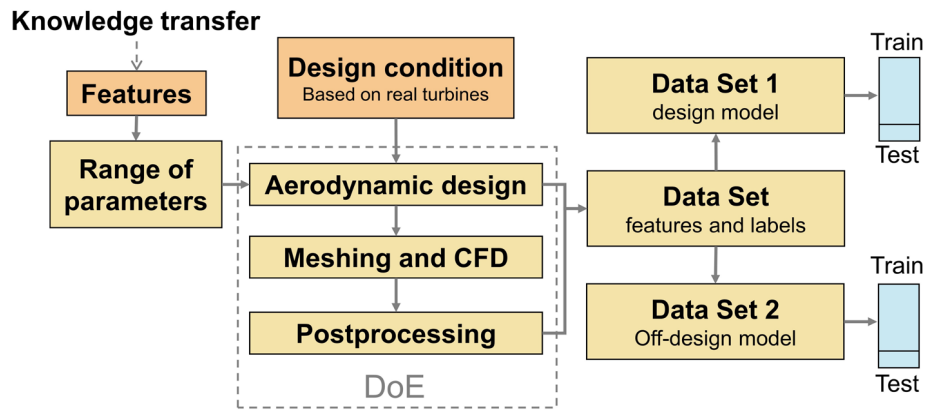

**Fig. 5.** Overview of the database construction process

2.3.1 Numerical method

A high-quality structural grid based on HOH type topology is generated by employing the commercial software Numeca. The computational grid and domain are shown in Fig. 6. The height of the first layer of the grid is $2\times10^{-4}$ mm, so the calculated $y+$ is about 1. There are about 1 million grid points in the stator channel and about 1.2 million grid points in the rotor channel, which is verified to satisfy the grid independence requirement. The length of the inlet section is about twice the axial chord length, and the length of the outlet section is more than three times the axial chord length.

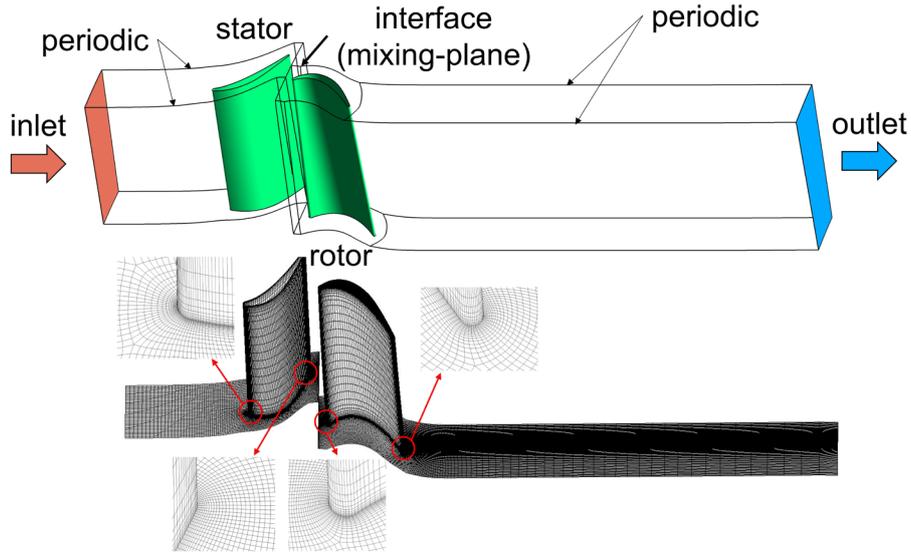

**Fig. 6.** Computational domain and grid for turbine cases

The numerical simulation is conducted using the commercial software ANSYS CFX to obtain the aerodynamic performance and flow in the helium turbine. The computational setups are as follows: helium ideal gas is used for the fluid model, the SST model is used for the turbulence model, and the $\gamma-\theta$ model is used for the transition model. The mixing-plane model is used between the stator and rotor blades. The calculations are considered to be converged when the residuals are lower than $1.0\times10^{-6}$. The numerical simulations are performed on a computational cluster, and the calculation of a single case takes about ten kernel hours.

2.3.2 Cases design process

Several sets of design input parameters are extended based on the design input parameters of an actual helium turbine as a benchmark. The range of the design input parameters is shown in Table 2. Based on these parameters, 102 sets of single-stage helium turbines are designed.

**Table 2** Range of design condition parameters in the database

| Parameters | Symbol | Range |
| --- | --- | --- |
| Expansion ratio | $\pi^*$ | 1.08-1.3 |
| Inlet total temperature(K) | $Tt_0$ | 800-1100 |
| Inlet total pressure(MPa) | $Pt_0$ | 6.25-20 |

The features of the model include one-dimensional and two-dimensional design parameters, covering most degrees of freedom in low-dimensional turbine design. Consequently, the following case design method is employed: A series of load coefficients, flow coefficients, reaction, and aspect ratios are selected at equidistant intervals. Within the corresponding parameter ranges, the blade profile parameters are chosen. The range of Zweifel coefficients in the test cases is set between 0.9 and 1.4. Loading distributions are divided into three forms: front-loaded, mid-loaded, and after-loaded, and are evenly assigned to all cases.

The one-dimensional design is conducted using a custom-built program. For a given set of input parameters, dimensionless parameters, including load coefficient, flow coefficient, reaction, and axial velocity ratio are used to calculate the velocity triangle and the meridian channel. The profile design and blade stacking are accomplished using the commercial software NREC. The Modified Pritchard parameterization method[43] is used for the profile generation. As for the blade stacking, a straight line is employed as the stacking curve. In terms of blade stacking, a straight line is utilized as the stacking curve, with the stator being stacked along the leading edge and the rotor stacked along the center of gravity.

The CFD method is utilized to obtain the performance of each helium turbine. In addition to the design point condition, non-design conditions are also computed, including 70%, 80%, 90%, 100%, and 110% of the rotation speed with five specific points at each speed line resulting in approximately 2500 sets of samples. The post-processing of the CFD results involves extracting the average total temperature and average total pressure at the inlet and outlet of each blade, as well as the average exit static pressure and average Mach number. Using Eq (11) and Eq (12), the total pressure loss coefficient and velocity loss coefficient are calculated. This processing facilitates the acquisition of various performance parameters and aerodynamic characteristics of every helium turbine case.

$$Y_p = \frac{Pt_1 - Pt_2}{Pt_2 - P_{s2}} \tag{12}$$

$$\Delta \Phi^2 = \frac{Ma_2^2}{\frac{2}{\kappa-1}\left[\left(1+\frac{\kappa-1}{2}Ma^2\right)\left(\frac{Pt_2}{Pt_1}\right)^{\frac{\kappa-1}{\kappa}} - 1\right]} \tag{13}$$

2.3.3 Database construction

From the design process and post-processing, the design parameters and flow parameters can be extracted as follows:

$$[\beta_1, \beta_2, i, H/C, \sigma, t_{LE}/C, t_{TE}/O, Ma_1, Ma_2, Re_1, Re_2, \gamma, t_{max}/C, \xi_{des}, \xi/\xi_{des}].$$

The combination of the above parameters forms a comprehensive database, with the range of each parameter illustrated in Table 3.

**Table 3** Range of parameters for the helium turbine database

| Parameters | Min value | Max value |
|---|---|---|
| $\beta_1$ | -0.559 | 1.046 |
| $\beta_2$ | 0.453 | 1.324 |
| $i$ | -0.480 | 0.255 |
| $H/C$ | 0.788 | 3.021 |
| $\sigma$ | 0.609 | 1.877 |
| $t_{LE}/C$ | 0.037 | 0.080 |
| $t_{TE}/O$ | 0.021 | 0.158 |
| $Ma_1$ | 0.090 | 0.396 |
| $Ma_2$ | 0.183 | 0.681 |
| $Re_1$ | $2.285 \times 10^5$ | $2.517 \times 10^6$ |
| $Re_2$ | $2.572 \times 10^5$ | $3.571 \times 10^6$ |
| $\gamma$ | 0.314 | 1.065 |
| $t_{max}/C$ | 0.08 | 0.23 |
| $\xi_{des}$ | 0.9492 | 0.9897 |
| $\xi/\xi_{des}$ | 0.9735 | 1.0040 |

The establishment of the design point loss model requires a dataset containing the following parameters:

$$[\beta_1, \beta_2, H/C, \sigma, t_{TE}/O, Ma_1, Ma_2, Re_1, Re_2, \gamma, t_{max}/C, \xi_{des}]$$

The dataset for establishing the off-design point loss model contains the following parameters:

$$[\beta_1, \beta_2, i, t_{LE}/C, \xi/\xi_{des}]$$

To ensure the accuracy and generalization performance of the model, it is necessary to divide the database into train and test sets carefully. The train set and the test set occupies proportions of 80% and 20%, respectively. It is advisable to include samples from various regions of the parameter space in the test set.

## 3. Results and discussions

This section begins by presenting the training outcomes and the precision of the model. Subsequently, a detailed analysis is conducted on the selection guidelines for low-dimensional design parameters of helium turbines. Finally, a conventional helium turbine and a contra-rotating helium turbine are designed and compared.

**3.1 New model and precision analysis**

3.1.1 Model at the design point

The model was evaluated using the test set every 50 epochs during the model training process. The losses of the model on both the train set and test set were measured and recorded, shown in Fig. 7. Both the train loss and test loss demonstrate favorable convergence, and the final losses exhibit minimal disparity, indicating a good generalization performance. Eq represents the results of the trailing edge loss model (14).

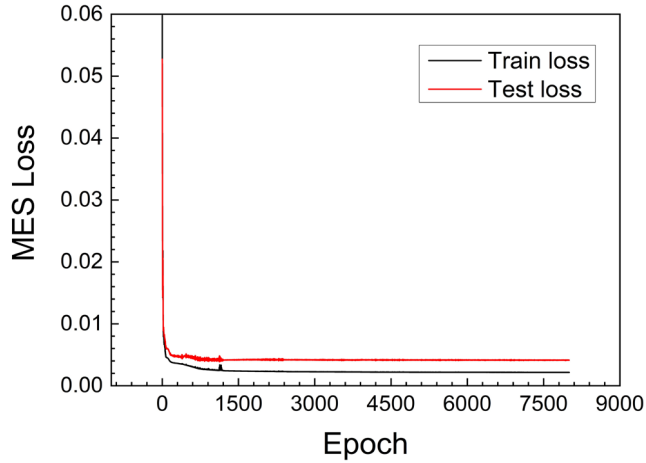

**Fig. 7.** Convergence history of MSE on train and test dataset

$$\Delta \Phi_{TET}^2 = \left( -0.2875 + 0.1069 \frac{\beta_1}{\beta_2} \left| \frac{\beta_1}{\beta_2} \right| \right) \left( \frac{t_{TE}}{O} \right)^3 \\ + \left( 0.7977 - 0.3601 \frac{\beta_1}{\beta_2} \left| \frac{\beta_1}{\beta_2} \right| \right) \left( \frac{t_{TE}}{O} \right)^2 + \left( 0.0812 - 0.0416 \frac{\beta_1}{\beta_2} \left| \frac{\beta_1}{\beta_2} \right| \right) \left( \frac{t_{TE}}{O} \right) \tag{14}$$

Model accuracy is evaluated using the test dataset, comparing the conventional gas turbine loss model with the new model proposed in this paper. The comparison between the predicted values and the CFD results, as well as the distribution of errors, is presented in Fig. 8. The predictions from the conventional model exhibit significant errors, with an overall error of 2.5%. In contrast, the new model demonstrates a high level of precision, with over 90% of the data points showing prediction errors within 0.5%.

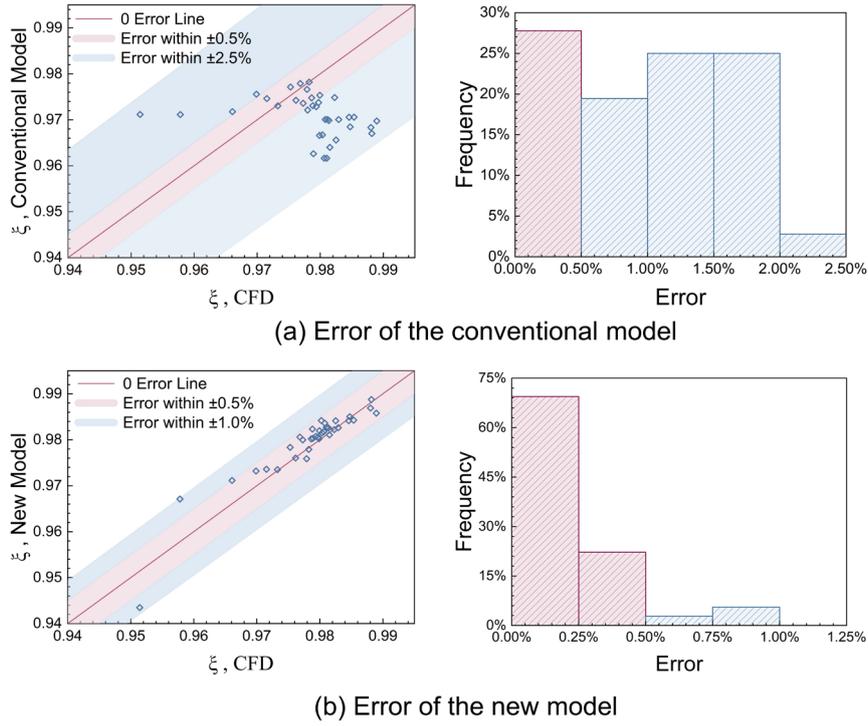

(a) Error of the conventional model

(b) Error of the new model

**Fig. 8.** Prediction error of the conventional model and the new model

3.1.2 Model at the off-design point

The Eq (15) and Eq (16) represent the calculation equation for the dimensionless incidence coefficient and the loss correlation obtained through the training process, respectively.

$$\chi = \frac{\alpha_1 - \alpha_{1(\text{des})}}{\left(\dfrac{\cos\beta_1}{\cos\beta_2}\right)^{1.76} (\beta_1 + \beta_2)\left(\dfrac{t_{TE}}{C}\right)^{1.12}} \quad (15)$$

$$\frac{\xi}{\xi_{\text{des}}} = -6.936\times10^{-5}\chi^4 - 7.073\times10^{-4}\chi^3 - 3.719\times10^{-3}\chi^2 - 4.917\times10^{-3}\chi + 1 \quad (16)$$

The performance of the off-design point loss model on the test dataset is illustrated in Fig. 9. Notably, approximately 98% of the data points show errors within ±0.5%, which suggests that the model achieves a high level of accuracy.

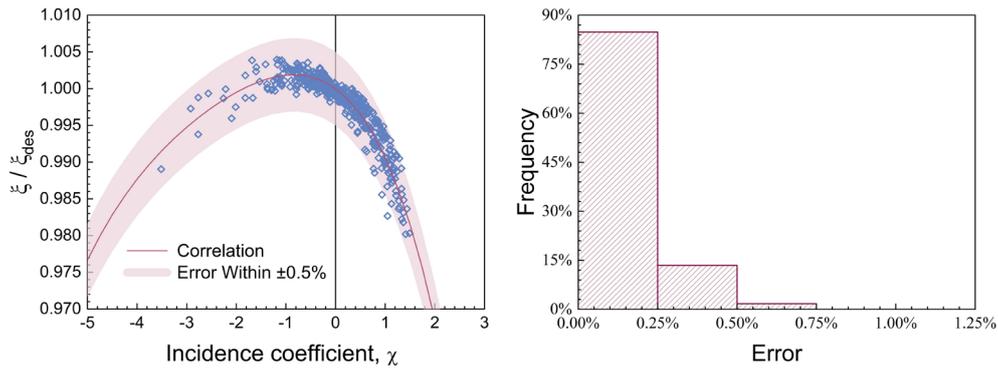

**Fig. 9.** Prediction error of the off-design point loss model

### 3.2 Design parameter selection analysis

The primary focus lies in examining the effect of helium properties on the selection of low-dimensional design parameters. The specific heat capacity of helium is approximately five times higher than that of air, which results in larger expansion work under the same expansion ratio. As shown in Fig. 10, at an expansion ratio of 1.3, the flow turning angle in the helium turbine has already surpassed 110 degrees, which will pose challenges to helium turbine design. Consequently, helium turbines typically employ a relatively smaller single-stage expansion ratio, leading to more stages.

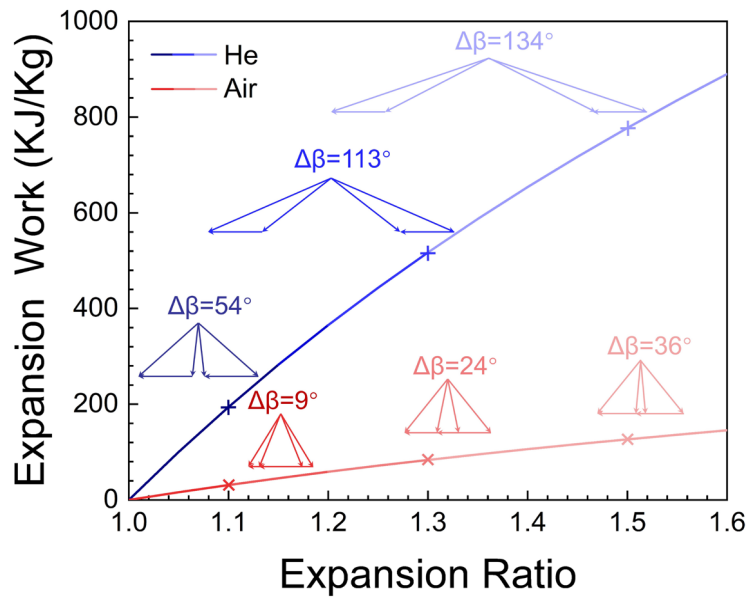

**Fig. 10.** Comparison of expansion work between helium and gas turbines

The low expansion ratio in helium turbines leads to a slight variation in density along the axial direction, resulting in a negligible change in the meridional plane and maintaining an axial velocity ratio of 1. The blade speed of each stage differs only slightly so that the flow coefficients of each stage are nearly equal. However, the situation differs for contra-rotating helium turbines with shorter axial lengths and relatively larger expansion ratios. The flow along the axial direction experiences a greater acceleration and a significant change in density. Resulting in a greater expansion of the meridian channel. Consequently, the axial velocity ratio and the flow coefficient can vary within a specific range.

During the design of turbines, the loading coefficient and flow coefficient are the key parameters, which are typically determined using the Smith chart [11]. A helium turbine Smith chart has been developed based on the established loss model to guide helium turbine design. Fig. 11 presents the Smith charts for gas and helium turbines, with the solid black line indicating the optimum relationship between the load and flow coefficients. The general trend of the helium turbine Smith chart is similar to that of gas turbines. However, the slope of the optimum line for helium turbines is greater, which means that the optimum loading coefficient for helium turbines is higher for the same flow coefficient. This is primarily due to the low Mach number in helium turbines under higher loads.

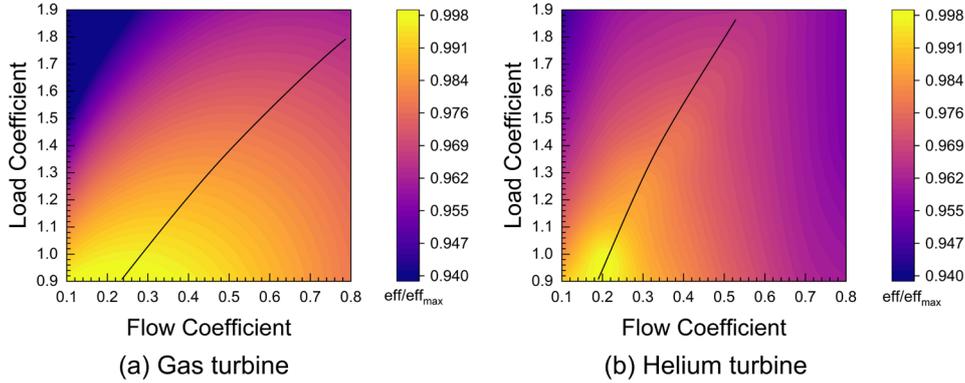

**Fig. 11.** Smith chart of gas turbines and helium turbines

The helium turbine loss model from [25] is employed to analyze the low-dimensional design parameters for the helium turbine in the GTHTR300 reactor[7], the design parameters of which are provided in Table 4. The correlation from [25] and the correlation based on the new model proposed in this paper are illustrated in Figure 11. For the actual loading coefficient of 1.407 in the GTHTR300, Fig. 12(a) demonstrates that the optimum flow coefficient is approximately 0.6. However, in Fig. 12(b), the optimum flow coefficient is around 0.4. Notably, this value is closer to the actual design value. This indicates that the model and design approach presented in this paper are more appropriate for helium turbines.

**Table 4** Design parameters of GTHTR300 helium turbine

| Parameters | Values |
|---|---|
| Mass flow rate(kg/s) | 441.8 |
| Shaft speed(rpm) | 3600 |
| Intel total temperature(K) | 1123 |
| Intel total pressure(MPa) | 6.88 |
| Mechanical power(MW) | 530 |
| Load coefficient | 1.407 |
| Flow coefficient | 0.434 |

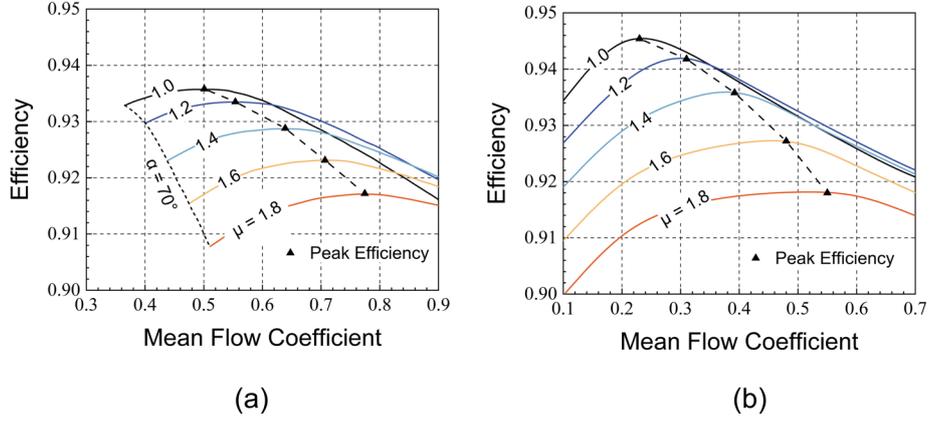

**Fig. 12.** Calculated efficiency of the helium turbine; (a) correlation from [25], (b) correlation derived from the new model

The contra-rotating helium turbine consists of multiple stages and requires certain design simplifications. Fig. 13 illustrates the velocity triangles of the contra-rotating helium turbine. The simplified approach involves setting the rotational speeds of the two axes to be equal. Due to the slight variation in the meridian plane, the blade speed of equivalent in each rotor. In the preliminary design, the axial velocity ratio is fixed at 1, ensuring the equality of flow coefficients. The Eq (17) expresses the relationship between the absolute outlet flow angle $\alpha_4$ of the second rotor stage and other design parameters.

$$\alpha_4 = \arctan \frac{\varphi \tan \alpha_1}{(\mu_2 - \mu_1)\tan \alpha_1 + \varphi} \tag{17}$$

When the load coefficient of the two rotor stages is equal, $\alpha_4$ is equal to $\alpha_1$. This means that the inlet velocity for the third stage is the same as that for the first stage, which significantly simplifies the design process. In practical design, the initial step follows the simplification approach, and modifications are made to the flow coefficients and axial velocity ratio based on the meridian plane.

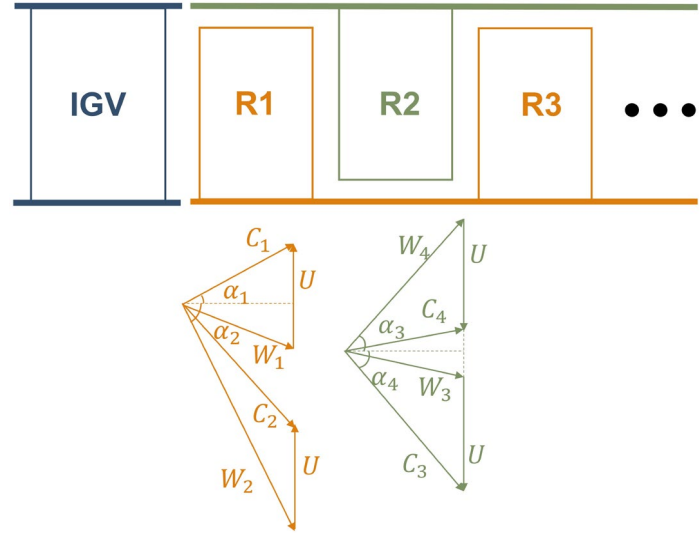

**Fig. 13.** Schematic of the contra-rotating helium turbine velocity triangle

In addition to the load coefficient and flow coefficient, the inlet flow angle is another fundamental parameter in the design of contra-rotating helium turbines, which affects the velocity triangles of both rotor stages. Fig. 14 illustrates the relationship between the efficiency of the two rotor stages and the inlet flow angle. The efficiency of the two rows of blades shows an opposite trend, indicating that a compromise should be made to achieve higher efficiency.

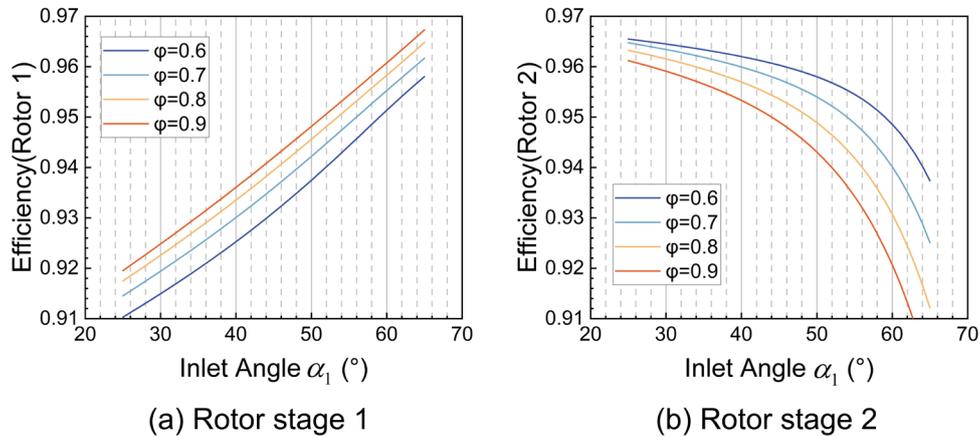

(a) Rotor stage 1         (b) Rotor stage 2

**Fig. 14.** Effect of the first stage inlet flow angle on the two rotor stages

### 3.3 Conventional and contra-rotating helium turbine

A highly loaded conventional helium turbine design and a highly loaded contra-rotating helium turbine design are conducted employing the loss model and low-dimensional parameter selection guidelines proposed in this paper. Subsequently, a comparison between the two types of designs is performed. The design requirements are presented in Table 5. The number of rotor rows is the same to ensure comparability between the two turbine designs. Additionally, both sets of turbines have identical rotational speeds and mean diameters.

Table 5 Design input parameters of helium turbine designs

| Input parameters | Values |
|---|---|
| Expansion ratio | 3.2 |
| Mass flow rate(kg/s) | 11.22 |
| RPM | 65000 |
| Intel total temperature(K) | 1100 |
| Intel total pressure (MPa) | 20 |

The conventional helium turbine design comprises four stages, with a single-stage expansion ratio of approximately 1.3, which is a highly loaded helium turbine design. The number of blades in each row is adjusted to attain a Zweifel coefficient of 1.1 at the mean radius, and the blade stacking curves are all straight. The design parameters are shown in Table 6. Visual representations of the design, including the meridional plane, blade profile at the mean radius, velocity triangle, and 3D blade diagram, are illustrated in Fig. 15.

Table 6 Design parameters of the conventional helium turbine design

| Design parameters | |
|---|---|
| Number of stages | 4 |
| Loading coefficient | 2.2 |
| Flow coefficient | 0.8 |
| Blades speed at mean radius (m/s) | 491 |
| Turning angle at mean radius (°) | 98 |
| Blade height of the first rotor (mm) | 8.7 |
| Zweifel coefficient | 1.1 |

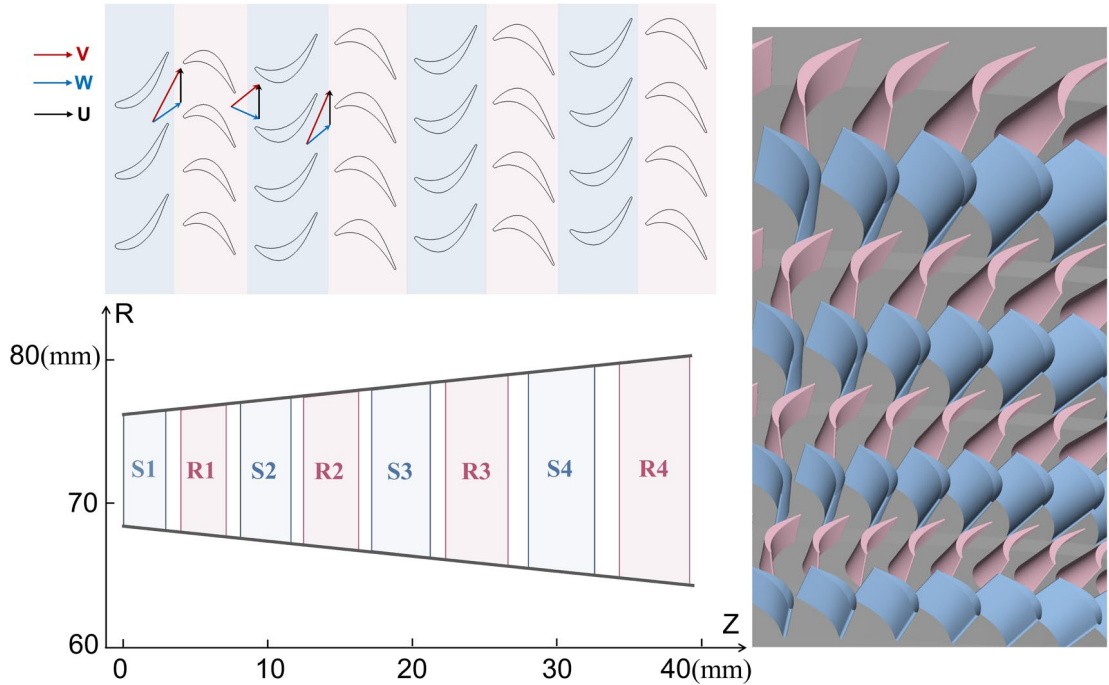

**Fig. 15.** The meridional plane, velocity triangle, profiles, and blades of the conventional helium turbine design

The contra-rotating helium turbine comprises inlet guide vanes followed by four rotor stages. As a highly loaded design, the individual rotor expansion ratio is also around 1.3. Due to the high expansion ratio and relatively short axial length, the angle of expansion in the meridional plane is relatively large. Therefore, adjustments are made to the axial velocity ratio and flow coefficient to control the degree of expansion. It should be noted that the expansion angle in the meridional plane for helium turbines with small expansion ratios is relatively small, and the adjustment is generally unnecessary. The Zweifel coefficient is set to 1.1 at the mean radius, and the blade stacking curves are straight. The design parameters are presented in Table 7. The meridional plane, blade profile at the mean radius, velocity triangle, and 3D blade diagram, are depicted in Fig. 16. The dashed lines in the figure represent the meridional before the adjustment of the axial velocity ratio. The inlet absolute flow angle is selected as 50°, allowing for relatively higher efficiency in both the odd and even stages. Notably, the flow turning angles for the first and third stages are only 54°, significantly reducing the axial length.

**Table 7** Design parameters of the contra-rotating helium turbine design

| Design parameters | |
|---|---|
| Number of rotors | 4 |
| Loading coefficient | 2.2 |
| Flow coefficient | 0.7 |
| Blades speed at mean radius (m/s) | 491 |
| Turning angle at mean radius (°) | 54/100 |
| Absolute intel angle (°) | 50 |
| Blade height of the first rotor (mm) | 10 |
| Zweifel coefficient | 1.1 |
| Number of blades in the first stage | 110 |

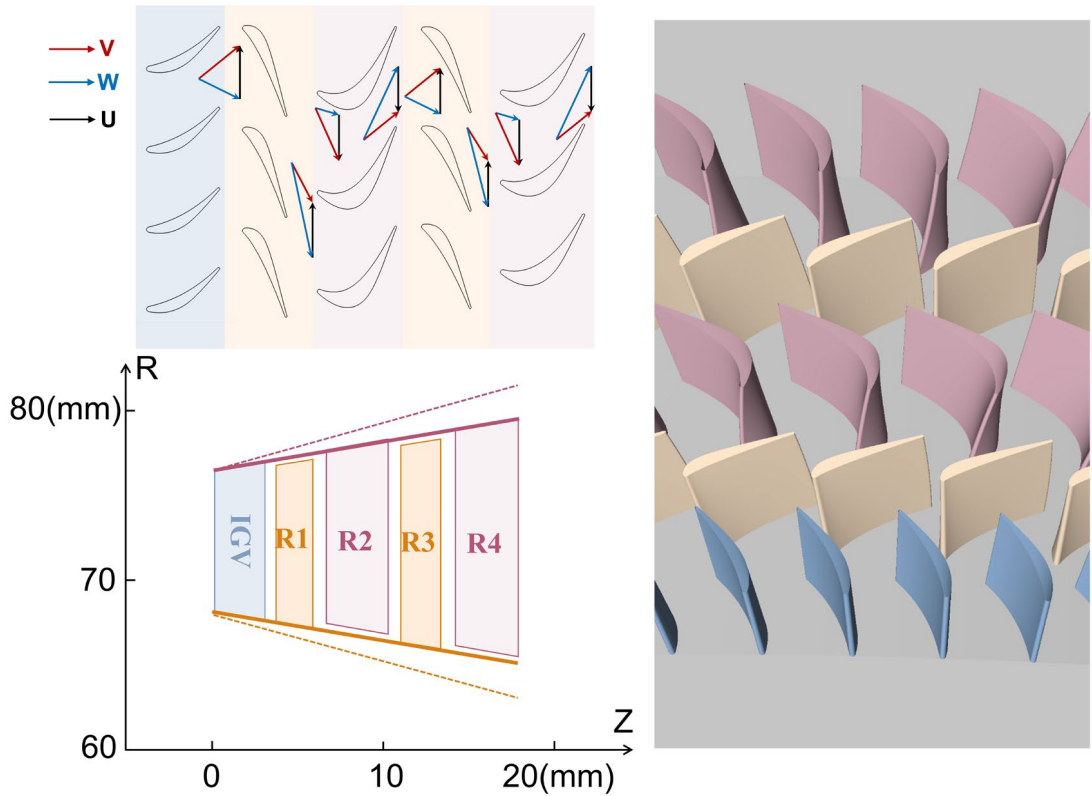

**Fig. 16.** The meridional, velocity triangle, profiles, and blades of the contra-rotating design

The results of the design and CFD calculation for the conventional and contra-rotating helium turbines are presented in Table 8. In terms of dimensions, the conventional design consists of 8 rows of blades, compared to 5 rows in the contra-rotating design. In addition, rotors with small turning angles in the contra-rotating turbine are short in axial length. Ultimately, these factors result in the axial dimension of the contra-rotating helium turbine being merely half that of the conventional helium turbine. In terms of performance, the contra-rotating helium turbine demonstrates a superior efficiency of 0.5% over the conventional turbine. This highlights the advantage of the contra-rotating turbine under highly loaded situations. However, the increased performance comes with the increased complexity in the structure. The arrangement of blades on two shafts raises challenges in terms of structure and strength. Consequently, the contra-rotating design is more suitable for size-sensitive applications, particularly in the aerospace industry. On the contrary, the conventional multi-stage helium turbine is a more reliable option in fields where size and weight considerations are not significant, such as nuclear reactor power generation.

Table 8 Results of the conventional and the contra-rotating design

| Parameters | Conventional | Contra-rotating |
| --- | --- | --- |
| Number of stages (rotor rows) | 4 | 4 |
| Length of turbine stages (mm) | 39.0 | 17.3 |
| Blades speed at mean radius (m/s) | 491 | 491 |
| Zweifel coefficient | 1.1 | 1.1 |
| Number of the first rotor blades | 126 | 110 |
| Total to total efficiency (%) | 89.07 | 89.51 |
| Mass flow rate (kg/s) | 11.32 | 11.31 |

## 4. Conclusions

This paper utilizes the knowledge transfer and Neural Network approach to establish a helium turbine loss model at design and off-design points. The developed loss model is then employed to derive the guidelines for low-dimensional design parameters selection for helium turbines, and a comparison is made against the gas turbine design experience. Furthermore, a contrastive analysis is conducted on the characteristics of a conventional multi-stage helium turbine design and a contra-rotating helium turbine design, with a single-stage expansion ratio of approximately 1.3. This paper draws the following conclusions:

(1) The helium turbine loss model, developed using the knowledge transfer and Neural Network approach, demonstrates remarkable accuracy at design and off-design points. The prediction errors of the model are below 0.5% at more than 90% of test samples surpassing the overall error of 2.5% shown in the gas turbine loss model. This result indicates that the model can provide sufficient guidance for the helium turbine design.

(2) The characteristics of helium turbines are pretty different from gas turbines. Helium turbines exhibit a smaller single-stage expansion ratio, an axial velocity ratio of 1, and approximately equal flow coefficients across each stage. The Smith charts derived for helium turbines reveal that, compared to gas turbines, the optimum flow coefficient for a given load coefficient is lower in helium turbines. For a 530MW helium turbine, the optimum flow coefficient calculated by the method proposed in this paper differs from the actual value by only 0.04. In contrast, the conventional gas turbine loss model yields a deviation of approximately 0.2. This suggests that the method in this paper outperforms the gas turbine experience in the design of helium turbines.

(3) A simplified method is proposed for the preliminary design of a contra-rotating helium turbine. To improve the efficiency of highly loaded contra-rotating turbines, controlling the considerable expansion in the meridian plane is necessary. This can be achieved by adjusting the axial velocity ratio and the flow coefficient. The influence of the inlet flow angle on the following two rotors exhibits a contrasting trend, so a compromise is required to achieve a higher overall efficiency.

(4) The contra-rotating design differs in several aspects from the conventional multi-stage design. The contra-rotating helium turbine features an axial length that is merely half that of the conventional helium turbine and offers an improved aerodynamic efficiency of 0.5%. However, the structure of the contra-rotating turbine is more complex. The characteristics inherent to these two designs determine that the contra-rotating helium turbine is more suitable for weight-sensitive applications, particularly aerospace. Conversely, the conventional multi-stage helium turbine is better suited for ground-based situations with lower loads but higher reliability requirements.

**Nomenclature**

| | |
|---|---|
| $X, Z$ | Feature variables |
| $A, B, C$ | Constant coefficient |
| $Y$ | Total pressure loss coefficient |
| $t_{max}$ | Maximum thickness of blades [mm] |
| $t_{TE}$ | Trailing edge thickness [mm] |
| $C$ | Blade chord [mm] |
| $O$ | Throat width [mm] |
| $H$ | Blade height [mm] |
| $i$ | Incidence angle [°] |
| $Re$ | Reynold number |
| $Ma$ | Mach number |
| $Pt$ | Total pressure [Pa] |
| $Ps$ | Static pressure [Pa] |
| $Tt$ | Total temperature [K] |
| $K$ | Correction factor |
| $N$ | Number of samples |
| $a, b, k, p, q$ | Undetermined coefficients |

*Greek symbols*

| | |
|---|---|
| $\alpha$ | Absolute flow angle [°] |
| $\beta$ | Relative flow angle [°] |
| $\sigma$ | Solidity |
| $\gamma$ | Blade stagger angle [°] |
| $\Delta\Phi^2$ | Energy loss coefficient |
| $\xi$ | Velocity loss coefficient |
| $\chi$ | Incidence coefficient |
| $\kappa$ | Specific heat ratio |
| $\mu$ | Load coefficient |
| $\varphi$ | Flow coefficient |

*Subscripts*

| | |
|---|---|
| 0 | Inlet of the turbine |
| 1 | Inlet of the blade |
| 2 | Outlet of the blade |
| 3 | Inlet of the second rotor in contra-rotating turbines |
| 4 | Outlet of the second rotor in contra-rotating turbines |
| G | Gas turbine model |
| H | Helium turbine model |
| $x$ | Axial component |
| $p$ | Profile loss |
| $s$ | Secondary loss |
| TET | |
| pred | Prediction value |
| des | Design point |
| inc | Incidence |

*Abbreviations*

| | |
|---|---|
| CFD | Computational fluid dynamic |
| IGV | Inlet guide vane |
| MSE | Mean square error |

**Acknowledgments**

This work was supported by the National Science and Technology Major Project [NO. J2019-II-0012-0032]. The supports are gratefully acknowledged.**References**

[1] Varvill R, Bond AC. The SKYLON Spaceplane - Progress to Realisation. J Br Interplanet Soc 2008;61:412–8.

[2] Jivraj F, Bond AC, Varvill R, Paniagua G. The Scimitar Precooled Mach 5 Engine, European Conference for Aero-Space Sciences 2007.

[3] Wang Z, Wang Y, Zhang J, Zhang B. Overview of the key technologies of combined cycle engine precooling systems and the advanced applications of micro-channel heat transfer. Aerosp Sci Technol 2014;39:31–9. https://doi.org/10.1016/j.ast.2014.08.008.

[4] McDonald CF. Helium turbomachinery operating experience from gas turbine power plants and test facilities. Appl Therm Eng 2012;44:108–42. https://doi.org/10.1016/j.applthermaleng.2012.02.041.

[5] No H, Kim JH, KIM H. A Review of Helium Gas Turbine Technology for High-temperature Gas-cooled Reactors. Nucl Eng Technol 2007;39. https://doi.org/10.5516/NET.2007.39.1.021.

[6] Herranz LE, Linares JI, Moratilla BY. Power cycle assessment of nuclear high temperature gas-cooled reactors. Appl Therm Eng 2009;29:1759–65. https://doi.org/10.1016/j.applthermaleng.2008.08.006.

[7] Takizuka T, Takada S, Yan X, Kosugiyama S, Katanishi S, Kunitomi K. R&D on the power conversion system for gas turbine high temperature reactors. Nucl Eng Des 2004;233:329–46. https://doi.org/10.1016/j.nucengdes.2004.08.017.

[8] Hong Y, Bi Y, Jing R, Ming Z, Ping Z, Qi Y, et al. Hong Y, Bi Y, Jing R, Ming Z, Ping Z, Qi Y, et al.. Construction of a 2 kW /4 K Helium Refrigerator for HT-7U. Plasma Sci Technol 2006;4:1305. https://doi.org/10.1088/1009-0630/4/3/007.

[9] Mito T, Sagara A, Imagawa S, Yamada S, Takahata K, Yanagi N, et al. Applied superconductivity and cryogenic research activities in NIFS. Fusion Eng Des 2006;81:2389–400. https://doi.org/10.1016/j.fusengdes.2006.07.086.

[10] Li X, Lv C, Yang S, Li J, Deng B, Li Q. Preliminary design and performance analysis of a radial inflow turbine for a large-scale helium cryogenic system. Energy 2019;167:106–16. https://doi.org/10.1016/j.energy.2018.10.179.

[11] Smith SF. A Simple Correlation of Turbine Efficiency. J R Aeronaut Soc 1965;69:467–70. https://doi.org/10.1017/S0001924000059108.

[12] Ainley DG, Mathieson G. A Method of Performance Estimation for Axial-Flow Turbines, 1951.

[13] Dunham J, Came PM. Improvements to the Ainley-Mathieson Method of Turbine Performance Prediction. J Eng Power 1970;92:252–6. https://doi.org/10.1115/1.3445349.

[14] Kacker SC, Okapuu U. A Mean Line Prediction Method for Axial Flow Turbine